\documentclass[onecolumn]{emulateapj}
\usepackage{apjfonts}

\begin{document}


\title{Constraints on the Self-Gravity of Radiation Pressure via Big Bang Nucleosynthesis}

\author{Saul Rappaport\altaffilmark{1},  Josiah Schwab\altaffilmark{1}, and Scott Burles\altaffilmark{1}}   

\altaffiltext{1}{Department of Physics and Kavli Institute for Astrophysics and Space Research, MIT, Cambridge, MA 02139; {\tt }}


\begin{abstract}
Using standard big-bang nucleosynthesis and present, high-precision measurements of light element abundances, we place constraints on the self-gravity of radiation pressure in the early universe. The self-gravity of pressure is strictly non-Newtonian, and thus the constraints we set are a direct test of this aspect of general relativity.
\end{abstract}



\section{Introduction}

Certain aspects of general relativity are well tested.  For example, the Schwarzschild metric has been quantitatively verified in the weak-field limit on small scales, e.g., the Solar system \citep{Shapiro64, Bertotti03} and binary radio pulsars \citetext{e.g., \citealp{Hulse75}; \citealp{Taylor79}; \citealp{Weisberg84}}; and on galaxy scales \citep[e.g.,][]{Bolton06}.  In another fundamental test of general relativity, the existence of gravity waves has been established  \citetext{e.g., \citealp{Taylor82}; \citealp{Weisberg84}, 2003}.  General relativity theory, in the form of the Friedmann--Robertson--Walker metric \citetext{\citealp{Friedmann22}, 1924; \citealp{Robertson35}, 1936; \citealp{Walker36}} and the Friedmann equations \citetext{\citealp{Friedmann22, Lemaitre27}}  which govern the expansion behavior of the Universe are used extensively in cosmology, and are at the core of this carefully and tightly woven paradigm.  However, it is probably fair to say that the Friedmann equations, while providing a very self-consistent and highly successful framework for cosmology, have not been subjected to extensive, independent testing.  In this paper we show that one particular aspect of general relativity, i.e., the self-gravity of pressure can be tested quantitatively.



The development of sophisticated big-bang nucleosynthesis (BBN) codes \citep{Wagoner67}, coupled with measurements of the relevant nuclear reaction rates \citep{Caughlan88}, have allowed observations of light elemental abundances to become powerful tools with which to investigate the evolution of the early universe. Computational predictions over a wide range of parameter space, when compared with observations, have yielded constraints on the current-epoch baryon density \citep{Wagoner73}, neutrino physics \citep{Yang79, Kawasaki94}, the fine structure constant \citep{Bergstrom99}, the gravitational constant \citep{Yang79, Copi04}, primordial magnetic fields \citep{Kernan96}, and other parameters of astrophysical interest. 

Increasingly exact measurements of elemental abundances, as well as augmented understanding of the processes (i.e., stellar nucleosynthesis) which have altered the original abundances, allow these restrictions to be continually refined. Deuterium abundances \citep{Black73, Omeara06}, helium abundances \citep{Peebles66, Izotov07}, and lithium abundances, including the effects of stellar mixing \citep{Pinsonneault02} have all been well measured. More recently, observations of the cosmic microwave background (CMB) have yielded an independent estimate of $\eta$, the baryon to photon ratio \citep{Spergel07}.


\section{Analysis}
\subsection{Friedmann Equations}

In the standard Friedmann--Robertson--Walker (FRW) cosmology, the expansion of the scale factor, $a$, of the universe as a function of its density and pressure is given by the Friedmann equations. In terms of $\ddot{a}$, the ``acceleration'' of the scale factor is given by:
\begin{equation}
\frac{\ddot{a}}{a} = - \frac{4\pi G}{3} \left(\rho + \frac{3P}{c^2}\right) ~~~.
\end{equation}
where $\rho c^2$ is the energy density and $P$ is the pressure.  This is an exact solution of the $\mathcal{G}_{rr}$ Einstein field equation for a homogenous and isotropic universe.  Note that the $3P$ term, implying the self-gravity of pressure, has no analog in Newtonian gravity. The other well-used form of the Friedmann equations:
\begin{equation}
\left(\frac{\dot{a}}{a}\right)^2 =  \frac{8\pi G}{3} \rho  ~~~,
\end{equation}
arises from the ${\bf \mathcal{G}}_{tt}$ component of the Einstein field equations (for a flat Universe).  For any fluid with a significant pressure, and given its appropriate equation of state, eq.~(1) can be integrated to yield eq.~(2) -- but only, of course, if the $3P$ term is included.

It could be argued that eq.~(2) is adequate to derive the evolution of the scale factor, and yet it depends only on $\rho$, and not explicitly on $P$.  Yet, we know that without the $3P$ term in eq.~(1), the constant of proportionality in front of $\rho$ in eq.~(2) will be incorrect, as long as pressure is significant. 
The Friedmann equations have been frequently derived from Newtonian theory, in the context of a {\em pressureless} ($P=0$) universe, both for scientific and strictly pedagogical reasons \citetext{e.g., \citealp{Milne34}; \citealp{Uzan01}; see also \citealp{Jordan}}. However, since the concept of self-gravity of pressure does not exist in Newtonian gravity, it follows that eqs.~(1) and (2) cannot be properly derived from any purely Newtonian construction \citep{Rainsford00}.  Finally we note that the Friedmann equation given by eq.~(2) (in which $P$ does not explicitly appear) formally ``knows'' about the form of eq.~(1) (which does contain $P$) via the Bianchi identity.

Consider now the radiation dominated epoch of the universe which prevailed for the first few thousand years after the big bang.  Let us start with the Friedmann equation in the form of eq.~(1).  From special relativistic considerations alone, we know that $P=1/3 \rho c^2$.  Therefore the $\ddot a$ equation can be written as:
\begin{equation}
\frac{\ddot{a}}{a} = - \frac{4\pi G(1+\chi)}{3} \rho ~~~,
\end{equation}
where $\chi$ is used as a placemarker for the pressure term (i.e., $\chi = 1$ if the pressure term is present, and $\chi = 0$ if not).  Since $\rho \propto a^{-4}$ for radiation and relativistic particles (e.g., neutrinos) undergoing an adiabatic expansion, eq.~(3) can be integrated to yield:
\begin{equation}
\left(\frac{\dot{a}}{a}\right)^2 ~=~~ \frac{8\pi G}{3} \left(\frac{1+\chi}{2} \right)\rho~~~,
\end{equation}
Variation of the value of $\chi$ away from unity would would significantly affect the outcome of BBN, degenerate with a variation in $G$ (the gravitational constant) or $N_{\nu}$ (the number of neutrino families). However, assuming these latter two to be constant at their nominal values of 1 and 3 respectively, nucleosynthetic constraints are able to place limits on $\chi$ and thus test the validity of this aspect of general relativity in this regime.

\subsection{Nucleosynthesis Calculations}

Nucleosynthesis calculations were performed with the canonical Kawano BBN codes \citep{Kawano92} which have been modified and updated with the latest reaction rates. The codes offer the ability to alter relevant early universe parameters such as $\eta$ and $G$. Since the parameter we seek to constrain, i.e., $(1+\chi$)/2, is multiplicative with $G$, we simply vary $G$ as a surrogate.  Altering the parameters over repeated runs of these codes yields abundances of light elements as a function of $\eta$ and $\chi$. 

In this work $\sim$75000 BBN calculations were performed, varying $\chi$ between $-1/2$ and 7 in steps of 0.02 (or equivalently $G/G_0$ from 1/4 to 4 in linear steps of 0.01) and varying $\log_{10}\eta$ from ${-10}$ to ${-9}$ in steps of 1/200 dex. 
The result of these calculations, is a grid of $\{\eta$, $\chi\}$ pairs, each point with a set of associated elemental abundances (see Fig.\,1).

As estimated in \citet{Burles01b} the uncertainties in the nuclear reaction rates leading to deuterium are 3.4\% (95\% confidence limit), and as estimated in \citet{Burles01a} $\gtrsim $0.1\% (95\% confidence limit) for He-4. 

\begin{figure*}[t]
\centering
\includegraphics[width = 5in]{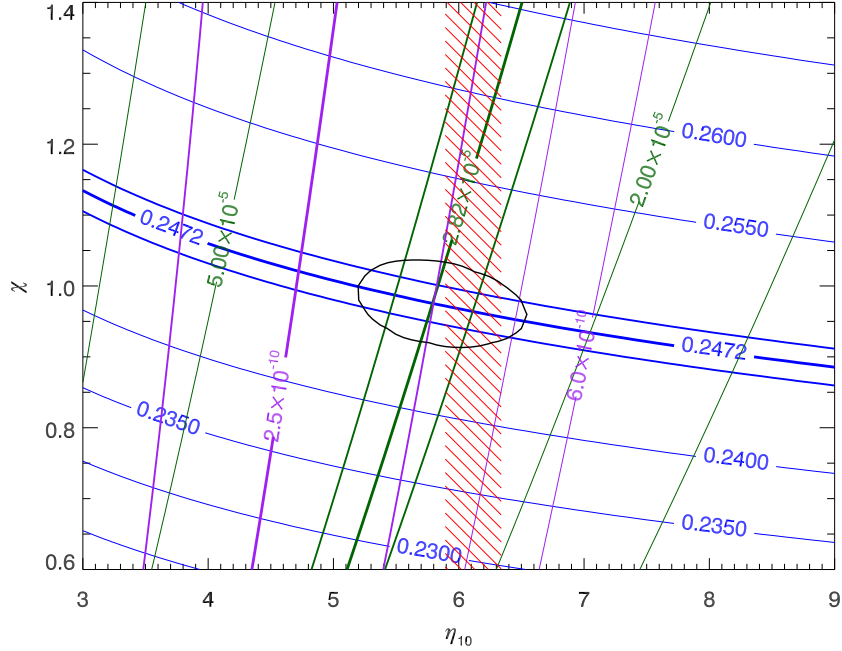}

\caption{Combination of the various constraints on $\chi$. The red-hatched region indicates the independent WMAP constraint on $\eta_{10}$ (i.e., the baryon/photon ratio in units of $10^{10}$). The bold contours represent the observed estimates and their $1\,\sigma$ errors. Green contours are deuterium abundances $(D/H)_p$. Blue contours are for the mass ratio of He-4 $(Y_p)$. Purple contours represent lithium abundances $(^7Li/H)_p$. The black ellipse is the 90-percent confidence contour based on the helium-4 and deuterium measurements.}

\end{figure*}

\newpage

\subsection{Light Element Observations}

Deuterium provides an excellent constraint because, as has been understood for the last few decades, the observed amounts necessitate that it must have formed in the big bang rather than in stellar or galactic processes \citep{Reeves73}. It is equally fortunate that it is extremely sensitive to the baryon to photon ratio, $\eta$ \citep{Burles01a}. Deuterium measurements from along the line of sight to high redshift quasars have culminated in the current determination of $\log_{10} (D/H)_p = -4.55 \pm 0.04$ \citetext{$1\,\sigma$ confidence; \citealp{Omeara06}}.  Contours of constant deuterium abundance are shown as green curves in Fig.\,1.


Lithium-7 was initially a frontrunner as a nucleosynthetic constraint. However it has somewhat fallen out of favor because of large uncertainties arising from the systematic effects in stellar depletion models. \citet{Pinsonneault02} determine the primordial lithium abundance as $\log_{10} (Li /H)_p = -9.6 \pm 0.2$ ($1\,\sigma$ confidence), which while consistent at the 1-$\sigma$ level with the estimates derived from other light elements, is unable to provide any additional significant constraint due to the relatively large error bars. Until this situation can be improved, lithium-7 will not likely play a strong role in constructing BBN-based parameter limits. Contours of constant lithium abundance are shown as purple curves in Fig.\,1.

Helium-4, though not as sensitive to $\eta$ as deuterium, also serves as a useful constraint on $\chi$. The primordial abundance is determined from present-day observations of HII regions, coupled with models of galactic chemical evolution. The primordial He-4 mass fraction, $Y_p$, is most recently reported to be $0.2472 \pm 0.0012$ \citetext{$1\,\sigma$ confidence; \citealp{Izotov07}}. Contours of constant helium-4 abundance are shown as blue curves in Fig.\,1.  

\subsection{WMAP Constraints}

Observations of the cosmic microwave background are able to provide an independent estimate of the baryon/photon ratio $\eta_{10}$ (i.e., $\eta$ in units of $10^{10}$). The {\em WMAP} collaboration \citep{Spergel07} gives as the three-year measurement $\eta_{10} =  6.116_{-0.249}^{+0.197}$. This measurement assumes $\chi = 1$, and therefore does not provide any additional constraint. However, as shown in Figure 1, the {\em WMAP} value of $\eta_{10}$ is in excellent agreement with the light element constraints.

\subsection{Constraints on $\chi$}

Figure 1 displays the results of our analysis. A section of the $\{\eta$, $\chi\}$ parameter space is shown, with {\em number} density (relative to hydrogen) contours for deuterium shown in green, and for {\em mass} fraction of He-4 shown in blue. The purple contours represent the Li-7 number density (relative to hydrogen).  The boldest contours indicate the current best abundance determinations, with the adjacent intermediate-width contours indicating the corresponding ranges of uncertainties ($1\,\sigma$ confidence).   

Assuming statisically-independent Gaussian errors, one can calculate the probability, via a maximum likelihood analysis, that the abundance determinations agree with the corresponding results of the BBN calculations at a given point in the $\{\eta$, $\chi\}$ parameter space. Only the constraints due to D and He-4 were used to compute the uncertainties in $\chi$ and $\eta$ (see \S 2.3 for an explanation of why Li-7 was not included).  The black ellipse is the 90\% probability contour.



\section{Conclusions}
As illustrated in Figure 1, the combined constraints correspond well with the general relativity prediction of $\chi = 1$ and the WMAP constraint on $\eta_{10}$. The 90-percent confidence contour gives a limit of $\chi = 0.97\pm 0.06$. Thus, we have used current light element observations and BBN computations as a test of the general relativistic self-gravity of pressure. The agreement is good to within $\sim$$6\%$.  We reiterate that our result can also be expressed as a 
constraint on the Newtonian constant, $G$ (during the BBN epoch), the number of light neutrino 
families, or degeneracies among neutrinos (see \S1 for references).  Assuming that these latter parameters take on their nominally expected values, then the self-gravity associated with the pressure of light and neutrinos during the BBN epoch has been quantitatively measured.



\acknowledgements 
JS acknowledges support from the Paul E. Gray (1954) Endowed Fund for UROP.  We thank Al Levine, Scott Hughes, and Ed Bertschinger for very helpful discussions.



\begin{thebibliography}{34}
\expandafter\ifx\csname natexlab\endcsname\relax\def\natexlab#1{#1}\fi

\bibitem[{Bergstr\"om {et~al.}(1999)Bergstr\"om, Iguri, \&
  Rubinstein}]{Bergstrom99}
Bergstr\"om, L., Iguri, S., \& Rubinstein, H. 1999, Phys. Rev. D, 60, 045005

\bibitem[{{Bertotti} {et~al.}(2003){Bertotti}, {Iess}, \&
  {Tortora}}]{Bertotti03}
{Bertotti}, B., {Iess}, L., \& {Tortora}, P. 2003, \nat, 425, 374

\bibitem[{{Black} \& {Dalgarno}(1973)}]{Black73}
{Black}, J.~H., \& {Dalgarno}, A. 1973, \apjl, 184, L101

\bibitem[{{Bolton}, {Rappaport}, \& {Burles}(2006)}]{Bolton06}
{Bolton}, A.~S., {Rappaport}, S., \& {Burles}, S. 2006, \prd, 74, 061501

\bibitem[{{Burles} {et~al.}(2001a){Burles}, {Nollett}, \& {Turner}}]{Burles01a}
{Burles}, S., {Nollett}, K.~M., \& {Turner}, M.~S. 2001a, \apjl, 552, L1

\bibitem[{Burles {et~al.}(2001b)Burles, Nollett, \& Turner}]{Burles01b}
Burles, S., Nollett, K.~M., \& Turner, M.~S. 2001b, Phys. Rev. D, 63, 063512

\bibitem[{{Caughlan} \& {Fowler}(1988)}]{Caughlan88}
{Caughlan}, G.~R., \& {Fowler}, W.~A. 1988, Atomic Data and Nuclear Data
  Tables, 40, 283

\bibitem[{{Coc} \& {Vangioni}(2005)}]{Coc05}
{Coc}, A., \& {Vangioni}, E. 2005, in From Lithium to Uranium: Elemental
  Tracers of Early Cosmic Evolution, ed. V.~{Hill}, P.~{Fran{\c c}ois}, \&
  F.~{Primas}, Vol. 228, 13--22

\bibitem[{{Copi} {et~al.}(2004){Copi}, {Davis}, \& {Krauss}}]{Copi04}
{Copi}, C.~J., {Davis}, A.~N., \& {Krauss}, L.~M. 2004, Physical Review
  Letters, 92, 171301

\bibitem[{{Friedmann}(1922)}]{Friedmann22}
{Friedmann}, A. 1922, Z. Phys, 10, 377

\bibitem[{{Friedmann}(1924)}]{Friedmann24}
{Friedmann}, A. 1924, Z. Phys, 21, 326

\bibitem[{{Hulse} \& {Taylor}(1975)}]{Hulse75}
{Hulse}, R.~A., \& {Taylor}, J.~H. 1975, \apjl, 195, L51

\bibitem[{{Izotov} {et~al.}(2007){Izotov}, {Thuan}, \&
  {Stasi{\'n}ska}}]{Izotov07}
{Izotov}, Y.~I., {Thuan}, T.~X., \& {Stasi{\'n}ska}, G. 2007, \apj, 662, 15

\bibitem[{{Jordan}(2005)}]{Jordan}
{Jordan}, T.F. 2005, Amer. J. Phys., 73, 653

\bibitem[{Kawano(1992)}]{Kawano92}
Kawano, L. 1992, FERMILAB-Pub-92/04-A

\bibitem[{{Kawasaki} {et~al.}(1994){Kawasaki}, {Kernan}, {Kang}, {Scherrer},
  {Steigman}, \& {Walker}}]{Kawasaki94}
{Kawasaki}, M., {Kernan}, P., {Kang}, H.-S., {Scherrer}, R.~J., {Steigman}, G.,
  \& {Walker}, T.~P. 1994, Nuclear Physics B, 419, 105

\bibitem[{Kernan {et~al.}(1996)Kernan, Starkman, \& Vachaspati}]{Kernan96}
Kernan, P.~J., Starkman, G.~D., \& Vachaspati, T. 1996, Phys. Rev. D, 54, 7207



\bibitem[{{Lema{\^i}tre}(1927)}]{Lemaitre27}
{Lema{\^i}tre}, G. 1927, Ann. Soc. Sci. Bruxelles, A47, 49.

\bibitem[{Milne \& McCrea(1934)}]{Milne34}
Milne, E., \& McCrea, W. 1934, QJ Maths, 5, 73


\bibitem[{{O'Meara} {et~al.}(2006){O'Meara}, {Burles}, {Prochaska}, {Prochter},
  {Bernstein}, \& {Burgess}}]{Omeara06}
{O'Meara}, J.~M., {Burles}, S., {Prochaska}, J.~X., {Prochter}, G.~E.,
  {Bernstein}, R.~A., \& {Burgess}, K.~M. 2006, \apjl, 649, L61

\bibitem[{Peebles(1966)}]{Peebles66}
Peebles, P. J.~E. 1966, Phys. Rev. Lett., 16, 410

\bibitem[{{Pinsonneault} {et~al.}(2002){Pinsonneault}, {Steigman}, {Walker}, \&
  {Narayanan}}]{Pinsonneault02}
{Pinsonneault}, M.~H., {Steigman}, G., {Walker}, T.~P., \& {Narayanan}, V.~K.
  2002, \apj, 574, 398

\bibitem[{{Rainsford}(2000)}]{Rainsford00}
{Rainsford}, T. 2000, General Relativity and Gravitation, 32, 719

\bibitem[{{Reeves} {et~al.}(1973){Reeves}, {Audouze}, {Fowler}, \&
  {Schramm}}]{Reeves73}
{Reeves}, H., {Audouze}, J., {Fowler}, W.~A., \& {Schramm}, D.~N. 1973, \apj,
  179, 909
  
 \bibitem[{{Robertson}(1935)}]{Robertson35}
{Robertson}, H.P. 1935, ApJ, 82, 284

\bibitem[{{Robertson}(1936)}]{Robertson36}
{Robertson}, H.P. 1936, ApJ, 83, 187

\bibitem[{{Shapiro}(1964)}]{Shapiro64}
{Shapiro}, I.~I. 1964, Physical Review Letters, 13, 789

\bibitem[{{Spergel} {et~al.}(2007){Spergel}, {Bean}, {Dor{\'e}}, {Nolta},
  {Bennett}, {Dunkley}, {Hinshaw}, {Jarosik}, {Komatsu}, {Page}, {Peiris},
  {Verde}, {Halpern}, {Hill}, {Kogut}, {Limon}, {Meyer}, {Odegard}, {Tucker},
  {Weiland}, {Wollack}, \& {Wright}}]{Spergel07}
{Spergel}, D.~N., {Bean}, R., {Dor{\'e}}, O., {Nolta}, M.~R., {Bennett}, C.~L.,
  {Dunkley}, J., {Hinshaw}, G., {Jarosik}, N., {Komatsu}, E., {Page}, L.,
  {Peiris}, H.~V., {Verde}, L., {Halpern}, M., {Hill}, R.~S., {Kogut}, A.,
  {Limon}, M., {Meyer}, S.~S., {Odegard}, N., {Tucker}, G.~S., {Weiland},
  J.~L., {Wollack}, E., \& {Wright}, E.~L. 2007, \apjs, 170, 377

\bibitem[{{Taylor} {et~al.}(1979){Taylor}, {Fowler}, \& {McCulloch}}]{Taylor79}
{Taylor}, J.~H., {Fowler}, L.~A., \& {McCulloch}, P.~M. 1979, \nat, 277, 437

\bibitem[{{Taylor} \& {Weisberg}(1982)}]{Taylor82}
{Taylor}, J.~H., \& {Weisberg}, J.~M. 1982, \apj, 253, 908

\bibitem[{{Uzan} \& {Lehoucq}(2001)}]{Uzan01}
{Uzan}, J.-P., \& {Lehoucq}, R. 2001, European Journal of Physics, 22, 371

\bibitem[{{Wagoner}(1973)}]{Wagoner73}
{Wagoner}, R.~V. 1973, \apj, 179, 343

\bibitem[{{Wagoner} {et~al.}(1967){Wagoner}, {Fowler}, \& {Hoyle}}]{Wagoner67}
{Wagoner}, R.~V., {Fowler}, W.~A., \& {Hoyle}, F. 1967, \apj, 148, 3

\bibitem[{{Walker}(1936)}]{Walker36}
{Walker}, A.G. 1936, Proc. Lon. Math. Soc., 42, 90

\bibitem[{{Weisberg} \& {Taylor}(1984)}]{Weisberg84}
{Weisberg}, J.~M., \& {Taylor}, J.~H. 1984, Physical Review Letters, 52, 1348

\bibitem[{{Weisberg} \& {Taylor}(2003)}]{Weisberg03}
{Weisberg}, J.~M., \& {Taylor}, J.~H. 2003, in Astronomical Society of the
  Pacific Conference Series, Vol. 302, Radio Pulsars, ed. M.~{Bailes}, D.~J.
  {Nice}, \& S.~E. {Thorsett}, 93

\bibitem[{{Yang} {et~al.}(1979){Yang}, {Schramm}, {Steigman}, \&
  {Rood}}]{Yang79}
{Yang}, J., {Schramm}, D.~N., {Steigman}, G., \& {Rood}, R.~T. 1979, \apj, 227,
  697

\end{thebibliography}



\end{document}